\documentclass[aps,
prd,
nofootinbib,
showpacs,
twocolumn,
superscriptaddress,
preprintnumbers,
floatfix]{revtex4-2}
\usepackage[bookmarks, linktocpage, colorlinks = true, linkcolor = blue, urlcolor  = blue, citecolor = blue, anchorcolor = green, hyperindex = true, hyperfigures]{hyperref}

\usepackage{graphicx} 
\usepackage{dcolumn}
\usepackage{multirow}
\usepackage{amsmath, amssymb}
\usepackage{slashed}
\usepackage[usenames]{color}
\usepackage{float}

\newcommand{\Th}{T_{\mathrm{H}}}

\newcommand{\phiB}{\phi_{B}}

\newcommand{\phitot}{\phi_{\mathrm{tot}}}

\begin{document}

\title{QCD thermodynamics through the crossover\\as a gas of confining strings with repulsive interactions}
\author{Yuki Fujimoto}
\affiliation{Department of Physics, Niigata University, Ikarashi, Niigata 950-2181, Japan}
\affiliation{RIKEN Center for Interdisciplinary Theoretical and Mathematical Sciences (iTHEMS), RIKEN, Wako 351-0198, Japan}
\author{Volodymyr Vovchenko}
\affiliation{Physics Department, University of Houston, 3507 Cullen Blvd, Houston, TX 77204, USA}

\date{\today}

\begin{abstract}
    We investigate QCD thermodynamics in the intermediate-temperature regime using a gas of confining strings supplemented by excluded volume (EV) corrections.
    In this description, the discrete hadron resonance spectrum is replaced by an exponentially growing open-string spectrum characterized by a Hagedorn temperature $\Th$, related to the confining string tension.
    With a common EV parameter shared by mesons and baryons, the model gives a good description of bulk lattice-QCD thermodynamics above the pseudocritical temperature.
    Among the values considered, $\Th \simeq 300\,\text{MeV}$ provides the best agreement, lower than the value $\Th \simeq 320$--$340\,\text{MeV}$ inferred from ideal-string-gas fits to the vacuum hadron spectrum.
    A simultaneous calibration of the mesonic and baryonic EV parameters gives $b_M \simeq 0.19\,\text{fm}^3$ and $b_B \simeq 0.73\,\text{fm}^3$.
    Within the present spectral ansatz, the larger baryonic EV parameter indicates that stronger effective suppression is required in the baryon sector, although noticeable tensions remain in several conserved-charge observables.
\end{abstract}

\maketitle

\section{Introduction}

In recent years, the temperature region above the chiral pseudo-critical temperature $T_c \simeq 156$~MeV has attracted considerable interest as a regime where the microscopic degrees of freedom of QCD matter are still not fully understood.
Although QCD thermodynamics shows a smooth crossover at vanishing baryon chemical potential~\cite{Aoki:2006we}, it remains nontrivial how the hadronic description below $T_c$ gives way to partonic degrees of freedom at higher temperatures.
This question is closely related to recent proposals of an intermediate regime between the hadronic phase and the quark-gluon plasma~\cite{Rohrhofer:2019qwq, Rohrhofer:2019qal,Glozman:2022lda,Cohen:2023hbq,Fujimoto:2025sxx,Hanada:2025rca}, such as the stringy fluid~\cite{Rohrhofer:2019qwq,Cohen:2023hbq} or the Spaghetti of Quarks with Glueballs (SQGB)~\cite{Fujimoto:2025sxx}, with the latter potentially connected to Quarkyonic matter in the denser regime~\cite{McLerran:2007qj,Fujimoto:2023mzy,Bluhm:2024uhj,Bluhm:2026hoj,Kojo:2026zfk}.
Such proposals are motivated in part by recent lattice studies indicating that confinement-related correlations can persist above $T_c$~\cite{Rohrhofer:2019qwq,Rohrhofer:2019qal,Petreczky:2021zmz,Bala:2021fkm,Lowdon:2022xcl,Aarts:2023nax,Bala:2023iqu,Bazavov:2023dci,Mickley:2024vkm,Bala:2025ilf} (see also refs.~\cite{Rothkopf:2019ipj,Marczenko:2025nhj,Marczenko:2026yme,Marczenko:2026drt,Aarts:2026kpq}).
These observations suggest that deconfinement need not be viewed as an abrupt disappearance of all confining dynamics, but rather as a gradual reorganization of the relevant degrees of freedom.
This motivates examining how far a thermodynamic description based on confining strings can be extended above the chiral crossover.

The hadron resonance gas (HRG) model provides a natural starting point for this investigation, as it successfully describes QCD thermodynamics below $T_c$.
Once the hadron spectrum is fixed, the ideal HRG model has no adjustable parameters.
The experimentally established hadron spectrum exhibits an approximately exponential growth in the cumulative number of states over the resonance region extending up to masses of about 2 GeV.
Such growth is a central feature of Hagedorn's statistical bootstrap model, in which hadronic fireballs are recursively composed of lighter fireballs~\cite{Hagedorn:1965st,Frautschi:1971ij}.
At large masses, the density of states takes the generic form $\rho_{\rm H}(m) \sim m^{-a} e^{m/\Th}$.
The resulting ideal-gas partition function fails to converge for $T > \Th$, leading to the interpretation of $\Th$ as a limiting temperature of hadronic matter.
The connection between an exponential hadronic spectrum and quark liberation was subsequently explored in Ref.~\cite{Cabibbo:1975ig}.

Following Hagedorn's original proposal~\cite{Hagedorn:1965st}, to describe strongly interacting matter, a number of works considered a gas of hadronic resonances with an exponentially rising mass spectrum~\cite{Hagedorn:1980cv,Rafelski:1980rk,Kapusta:1982qd} characterized by a Hagedorn temperature $\Th \simeq 150$--$180$~MeV~\cite{Broniowski:2004yh,Noronha-Hostler:2009wof,Cleymans:2011fx,Lo:2015cca}, on the order of $T_c$.
In phenomenological applications, however, the power-law prefactor and normalization of the Hagedorn spectrum are model dependent, and the extracted value of $\Th$ consequently depends on the adopted parametrization and fitting procedure~\cite{Cohen:2011cr}.

Relativistic strings provide an alternative microscopic basis for the exponential spectrum~\cite{Fubini:1969qb,Huang:1970iq,Green:1987sp}, different from that of the statistical bootstrap model.
For open bosonic strings in $3+1$ dimensions, counting the physical transverse modes gives the asymptotic density of states $\rho_{\rm str}(m)\sim m^{-3/2}e^{m/T_{\rm H}}$~\cite{Fujimoto:2025sxx,Green:1987sp}.
The power-law prefactor is thus $m^{-3/2}$, rather than the $m^{-5/2}$ commonly used in phenomenological Hagedorn parametrization.
Within this string description, $\Th$ is related to the confining string tension $\sigma$ by $\Th = \sqrt{3 \sigma/(2\pi)}$, thereby linking the spectral growth to the confinement scale.
For the commonly used value $\sigma \simeq 1~\text{GeV/fm}$, this relation gives $\Th \sim 0.3$ GeV.
In fact, fits of the open-string spectrum to the observed hadron spectra yield a value of approximately $0.3$~GeV~\cite{Marczenko:2025nhj,Fujimoto:2025trl}, twice as large as $T_c$.

Meyer demonstrated that thermodynamics of pure Yang-Mills theory in the confined phase can accurately be described by a glueball resonance gas supplemented with an exponential spectrum of \emph{closed} bosonic strings~\cite{Meyer:2009tq}.
The corresponding open-string description of mesonic excitations from the same underlying idea was applied to full QCD thermodynamics in Ref.~\cite{Fujimoto:2025sxx}. 
The resulting equation of state was found to agree well with lattice-QCD data.
The construction was subsequently extended to baryons by treating them as quark-diquark systems~\cite{Fujimoto:2025trl}, motivated by Regge phenomenology~\cite{Selem:2006nd}.

With the single-string spectrum specified, we next consider residual interactions among strings.
Excluded-volume (EV) corrections have been employed to address the thermodynamic singularity associated with the Hagedorn spectrum~\cite{Hagedorn:1980kb,Gorenstein:1981fa,Kapusta:1982qd,Hagedorn:1982qh}.
It was later shown that a Hagedorn gas with a bag model type EV correction can be made qualitatively consistent with lattice QCD thermodynamics~\cite{Ferroni:2008ej,Vovchenko:2018eod}.
The EV correction has since become a common ingredient of the HRG model~\cite{Andronic:2012ut}, particularly to account for the short-range repulsion among baryons~\cite{Vovchenko:2016rkn}.
It improves the agreement with the lattice data in the vicinity of and even above $T_c$~\cite{Vovchenko:2014pka,Vovchenko:2017xad} and can affect the extraction of chemical freeze-out parameters in heavy-ion collisions~\cite{Vovchenko:2016ebv,Poberezhnyuk:2019pxs}.
Further, it is a viable mechanism for generating the rapid stiffening of dense matter in neutron stars~\cite{Fujimoto:2021dvn,Vovchenko:2026ubv} and can facilitate the dynamical generation of momentum shell structure in some implementations of quarkyonic matter~\cite{Jeong:2019lhv,Moss:2024uam}.

Here we incorporate the EV correction into the recently proposed description of the HRG as a gas of confining strings~\cite{Fujimoto:2025sxx,Marczenko:2025nhj}.
With a single EV parameter, the model yields an excellent description of lattice QCD thermodynamics up to $T \simeq 200$--$230$~MeV, considerably above $T_c$.
Among the values considered, $\Th \simeq 300$~MeV gives the best agreement with lattice thermodynamics, lower than the  $\Th \sim 340$~MeV~\cite{Fujimoto:2025trl} value favored by the vacuum hadron-spectrum fit, and toward the lower end of the $\sim 0.30$--$0.34$~GeV range associated with zero-temperature string-tension estimates~\cite{Marczenko:2025nhj}.
An analysis of conserved-charge susceptibilities further indicates that the baryon sector requires a larger EV parameter than the mesonic one, indicating that stronger effective suppression is required in the baryon sector, consistent with the recent observation of Ref.~\cite{Marczenko:2026yme} that an ideal string-gas description of lattice fluctuations underestimates the repulsion in the light-baryon sector.

\section{String gas with excluded-volume corrections}

Our model combines a string-based description of the hadronic spectrum with van der Waals-type excluded-volume (EV) corrections that account for residual interactions.

\subsection{Hadron resonance gas and string density of states}
In the Maxwell-Boltzmann approximation, a noninteracting particle species of mass $m$ contributes the following pressure per internal degree of freedom:
\begin{align}
    P(m) = \frac{m^2T^2}{2\pi^2} K_2 \left(\frac{m}{T}\right)\,,
\end{align}
where $K_2$ is the modified Bessel function of the second kind.
Integrating this contribution over the mass density of states gives the total pressure:
\begin{align}
    P = \int_0^\infty dm\, \rho(m) P(m)\,,
\end{align}
where $\rho(m)$ denotes the mass density of states.
In the HRG model, the mass spectrum is given by
\begin{align}
    \rho(m) = \sum_i d_i \delta(m - m_i)\,.
\end{align}
Here, $d_i$ and $m_i$ denote the degeneracy and mass of species $i$, respectively, and $i$ runs over all hadrons, including resonances and antiparticles.
We neglect resonance widths and treat each state as a stable particle.

Thus, the thermodynamics of the HRG model is determined entirely by the hadronic spectrum.
Recent studies have shown that the cumulative hadron spectrum is well described by an open-string model~\cite{Marczenko:2025nhj,Fujimoto:2025trl}.
The asymptotic density of states for open bosonic strings in $3+1$ dimensions is
\begin{align}
    \rho_{\rm str}(m) = \frac{\sqrt{2\pi}}{6\Th} \left(\frac{m}{\Th}\right)^{-3/2} e^{m/\Th}\,,
    \label{eq:openstring}
\end{align}
where $\Th$ is the Hagedorn temperature and is related to the tension of the confining string $\sigma$ by
\begin{align}
    \Th = \sqrt{\frac{3\sigma}{2\pi}} = \frac{1}{2\pi}\sqrt{\frac{3}{\alpha'}}\,,
    \label{eq:THopen}
\end{align}
where $\alpha' = 1/(2\pi \sigma)$ is the Regge slope.
Although this spectrum resembles the Hagedorn spectrum of the statistical bootstrap model, it has a distinct microscopic origin:
it follows from the quantization of transverse string modes.
Quantitatively, its power-law prefactor has exponent $-3/2$, rather than the $-5/2$ commonly used in empirical Hagedorn parametrization (see, e.g., Refs.~\cite{Broniowski:2000bj,Rafelski:2015xej}).
Huang and Weinberg obtained the latter exponent by counting all $D$ oscillator degrees of freedom~\cite{Huang:1970iq}, whereas counting only the $D-2$ physical transverse polarizations yields the correct exponent $-3/2$ in $D=4$.
Using this string spectrum, we approximate the hadron mass spectrum as
\begin{align}
    \rho(m) &\approx \sum_{i = \pi, K, \eta} d_{i} \delta(m - m_{i}) 
     + d_{\rm str}(m) \rho_{\rm str}(m)\,,
\end{align}
where $d_{\rm str}(m)$ is the degeneracy factor for the string density of states $\rho_{\rm str}(m)$~\eqref{eq:openstring}.
We retain the light pseudo-Nambu-Goldstone (NG) mesons explicitly in the first term\footnote{We note that the sum $\sum_{i=\pi,K,\eta}$ implicitly resolves the pseudo-NG $\pi$ and $K$ multiplets into $B$, $Q$, $S$ states necessary for the calculation of conserved charge susceptibilities.}, while remaining hadronic excitations are represented by the continuum string density \eqref{eq:openstring} above channel-dependent thresholds.
In this representation, a meson is modeled as an open confining string ending on a quark and an antiquark, whereas a baryon is modeled as a string ending on a quark and a diquark.
The mass-dependent degeneracy factor $d_{\rm str}(m)$ is
\begin{align}
    d_{\rm str}(m) &= \sum_i d_{\mathrm{str}, i}\theta(m - m_{\mathrm{str},i})\,.
\end{align}
Here, $i$ labels the string channels, and the corresponding degeneracies $d_{\mathrm{str},i}$ and thresholds $m_{\mathrm{str},i}$ are listed in Table~\ref{tab:string}.
Each row represents a channel with fixed values of $B$, $Q$, and $S$.
The flavor multiplicities of a given endpoint configuration are resolved into separate charge channels, while $d_{\mathrm{str},i}$ denotes the remaining spin multiplicity of the endpoint.
\begin{table}[t]
\caption{Threshold masses, quantum numbers, and degeneracies of the mesonic ($B=0$) and baryonic ($B=1$) string channels, following Ref.~\cite{Fujimoto:2025trl}.
Here, $l$ denotes a light quark ($u$ or $d$), and $[qq]$ and $(qq)$ denote spin-0 (``good'') and spin-1 (``bad'') diquarks, respectively.
Hadron labels in parentheses identify the reference states used to set the thresholds $m_{\mathrm{str},i}$.
Antibaryon channels, not shown explicitly, have the same thresholds and degeneracies, with the signs of $B$, $Q$, and $S$ reversed.}
\label{tab:string}
\begin{ruledtabular}
\begin{tabular}{lccccc}
Channel $i$
  & $m_{\mathrm{str},i}$ [GeV]
  & $I$ & $Q$ & $S$
  & $d_{\mathrm{str},i}$ \\
\colrule

\multicolumn{6}{c}{Mesonic channels ($B=0$)} \\[2pt]

$l\!-\!\bar l$ ($\rho$)
  & 0.775 & 0 & 0 & 0 & 4 \\
  &       & 1 & $-1$ & 0 & 4 \\
  &       & 1 & 0 & 0 & 4 \\
  &       & 1 & 1 & 0 & 4 \\[2pt]

$l\!-\!\bar s$ ($K^*$)
  & 0.896 & $1/2$ & 0 & 1 & 4 \\
  &       & $1/2$ & 1 & 1 & 4 \\[2pt]

$s\!-\!\bar l$ ($\bar K^*$)
  & 0.896 & $1/2$ & $-1$ & $-1$ & 4 \\
  &       & $1/2$ & 0 & $-1$ & 4 \\[2pt]

$s\!-\!\bar s$ ($\phi$)
  & 1.019 & 0 & 0 & 0 & 4 \\[5pt]

\multicolumn{6}{c}{Baryonic channels ($B=1$)} \\[2pt]

$l\!-\![ll]$ ($N$)
  & 0.938 & $1/2$ & 0 & 0 & 2 \\
  &       & $1/2$ & 1 & 0 & 2 \\[2pt]

$s\!-\![ll]$ ($\Lambda$)
  & 1.116 & 0 & 0 & $-1$ & 2 \\[2pt]

$l\!-\![ls]$ ($\Sigma$)
  & 1.189 & 1 & $-1$ & $-1$ & 2 \\
  &       & 1 & 0 & $-1$ & 2 \\
  &       & 1 & 1 & $-1$ & 2 \\[2pt]

$l\!-\!(ll)$ ($\Delta$)
  & 1.232 & $3/2$ & $-1$ & 0 & 4 \\
  &       & $3/2$ & 0 & 0 & 4 \\
  &       & $3/2$ & 1 & 0 & 4 \\
  &       & $3/2$ & 2 & 0 & 4 \\[2pt]

$s\!-\![ls]$ ($\Xi$)
  & 1.315 & $1/2$ & $-1$ & $-2$ & 2 \\
  &       & $1/2$ & 0 & $-2$ & 2 \\[2pt]

$s\!-\!(ss)$ ($\Omega$)
  & 1.672 & 0 & $-1$ & $-3$ & 4
\end{tabular}
\end{ruledtabular}
\end{table}

We note that Eq.~\eqref{eq:openstring} is an asymptotic expression.
Its application immediately above the channel-dependent thresholds is therefore a modeling assumption rather than a controlled large-mass approximation.
The thresholds anchor the continuum description to the lightest non-NG states in each channel, but the treatment of the transition from the discrete low-mass spectrum to the asymptotic string continuum remains a source of systematic uncertainty.

In principle, glueballs, modeled as closed strings~\cite{Meyer:2009tq}, also contribute.
Their thermodynamic contribution, however, is orders of magnitude smaller than that of open strings.
This suppression follows from the asymptotic density of states for closed bosonic strings:
\begin{align}
    \rho_{\rm closed}(m) = \frac{(2\pi)^3}{27\Th} \left(\frac{m}{\Th}\right)^{-4} e^{m/\Th}\,.
\end{align}
Here, we note that the relation between $\sigma$ and $\Th$ remains the same as in the open string case \eqref{eq:THopen}.
The more negative power-law exponent, $-4$ rather than $-3/2$ in Eq.~\eqref{eq:openstring}, strongly suppresses the closed-string contribution.
This can be seen in Fig. 1 in Ref.~\cite{Fujimoto:2025sxx}.
Accordingly, we neglect closed strings in the following analysis.

\subsection{Excluded-volume prescription}

We incorporate repulsive interactions using an EV prescription.
In the Boltzmann approximation, the ideal-gas partial pressure of channel $i$ is
\begin{align}
    P_i^{\mathrm{id}}(T,\mu_i) &= T\phi_i(T)e^{\mu_i/T}\,,\notag\\
    \mu_i&=B_i\mu_B+Q_i\mu_Q+S_i\mu_S\,.
\end{align}
Here, $i$ labels channels with definite conserved charges, and $\phi_i(T)$ denotes their ideal-gas number density at vanishing chemical potentials.
For a discrete particle species $i$ with mass $m_i$ and degeneracy $d_i$, $\phi_i(T)$ is
\begin{equation}
  \phi_i(T)
  = \frac{d_i m_i^2 T}{2\pi^2}
    K_2\!\left(\frac{m_i}{T}\right)\,.
\end{equation}
For a string channel $i$, the discrete mass is replaced by the
continuous spectral density $\rho_{\mathrm{str}}(m)$ above the
threshold $m_{\mathrm{str},i}$, giving
\begin{equation}
  \phi_i(T)
  = d_{\mathrm{str},i}
    \int_{m_{\mathrm{str},i}}^\infty\!
    dm\,\rho_{\mathrm{str}}(m) \frac{m^2 T}{2\pi^2}
    K_2\!\left(\frac{m}{T}\right)\,.
\end{equation}

We first consider a common-eigenvolume model in which all channels are assigned the same EV parameter $b$.
The prescription is defined by replacing the volume $V$ in the ideal-gas Helmholtz free energy with the available volume $V_{\mathrm{av}}$:
\begin{align}
    F(T,V,\{N_i\})
    &=
    \sum_i F_i^{\mathrm{id}}(T,V_{\mathrm{av}},N_i)\,, \notag\\
    V_{\mathrm{av}} &=V-b\sum_jN_j\,.
\end{align}
In the thermodynamic limit, differentiation with respect to $V$ and $N_i$ gives the effective chemical potentials $\mu_i^*=\mu_i-bP$.
Consequently, the grand-canonical pressure satisfies the following transcendental equation~\cite{Rischke:1991ke}:
\begin{align}
    P(T,\boldsymbol{\mu})
    &=
    \sum_i P_i^{\mathrm{id}}(T,\mu_i-bP) \notag \\
    &=
    T e^{-bP/T}\sum_i\phi_i(T)e^{\mu_i/T}\,,
    \label{eq:PEV}
\end{align}
where $\boldsymbol{\mu}=(\mu_B,\mu_Q,\mu_S)$.
Solving this self-consistent equation for $P(T,\boldsymbol{\mu})$ yields the EV-corrected pressure.
At zero chemical potentials, this reduces to
\begin{equation}
    P(T,\boldsymbol{0})
    =
    T\sum_i\phi_i(T)\,e^{-bP(T,\boldsymbol{0})/T}.    
\end{equation}
The ideal-gas pressure is recovered for $b=0$.

To allow different repulsive-core strengths in the mesonic and baryonic sectors, we also consider the multi-component non-diagonal EV prescription of Ref.~\cite{Vovchenko:2016ebv}.
The available volume now depends on the channel:
\begin{align}
    F(T,V,\{N_i\})
    &=
    \sum_i F_i^{\mathrm{id}}(T,V_i,N_i)\,,\notag \\
    V_i &= V-\sum_j\widetilde b_{ji}N_j\,.
\end{align}
The same thermodynamic derivatives yield coupled equations for the partial pressures,
\begin{align}
    P_i
    &=
    P_i^{\mathrm{id}}\!\left(T,\mu_i-\sum_j\widetilde b_{ij}P_j\right) \notag \\
    &=
    T\phi_i(T)
    \exp\!\left[\frac{\mu_i-\sum_j\widetilde b_{ij}P_j}{T}\right]\,,
    \label{eq:presmulti}
\end{align}
and the total pressure is given as the sum of $P_i$:
\begin{equation}
    P=\sum_iP_i\,.
\end{equation}
We adopt the coefficients
\begin{equation}
    b_{ij}=\frac{2\pi}{3}(r_i+r_j)^3\,,
    \qquad
    \widetilde b_{ij}
    =
    \frac{2b_{ii}b_{ij}}{b_{ii}+b_{jj}}\,,    
\end{equation}
which reproduce the second virial coefficients of a hard-sphere mixture~\cite{Gorenstein:1999ce}.
The effective radii $r_i$ parametrize the repulsive interactions.
We assign a constant radius $r_M$ to all mesonic channels and $r_B$ to baryonic and antibaryonic channels, and denote the corresponding EV parameters by
\begin{align}
    b_M=\frac{16\pi}{3}r_M^3\,,
    \qquad
    b_B=\frac{16\pi}{3}r_B^3\,.
    \label{eq:bMbB}
\end{align}
For $r_M=r_B$, all $\widetilde b_{ij}$ become equal and the common-$b$ prescription is recovered.
For unequal radii, we solve the coupled pressure equations numerically.
Thermodynamic observables follow from derivatives of the total pressure.

In this prescription, baryons repel both the other baryons and antibaryons with the same strength, and repel mesons with a reduced strength.
Some other implementations neglect repulsion in the meson--baryon and baryon--antibaryon sectors~\cite{Vovchenko:2016rkn}, which would require recalibrating the fitted parameters.

\section{Bulk thermodynamics}

We compare the bulk thermodynamics of the ideal and interacting string gases with lattice QCD at vanishing chemical potentials.
The common EV parameter is fitted to the pressure and then used to calculate the other thermodynamic observables.

\subsection{Ideal string gas}
\begin{figure*}
    \centering
    \includegraphics[width=0.96\textwidth]{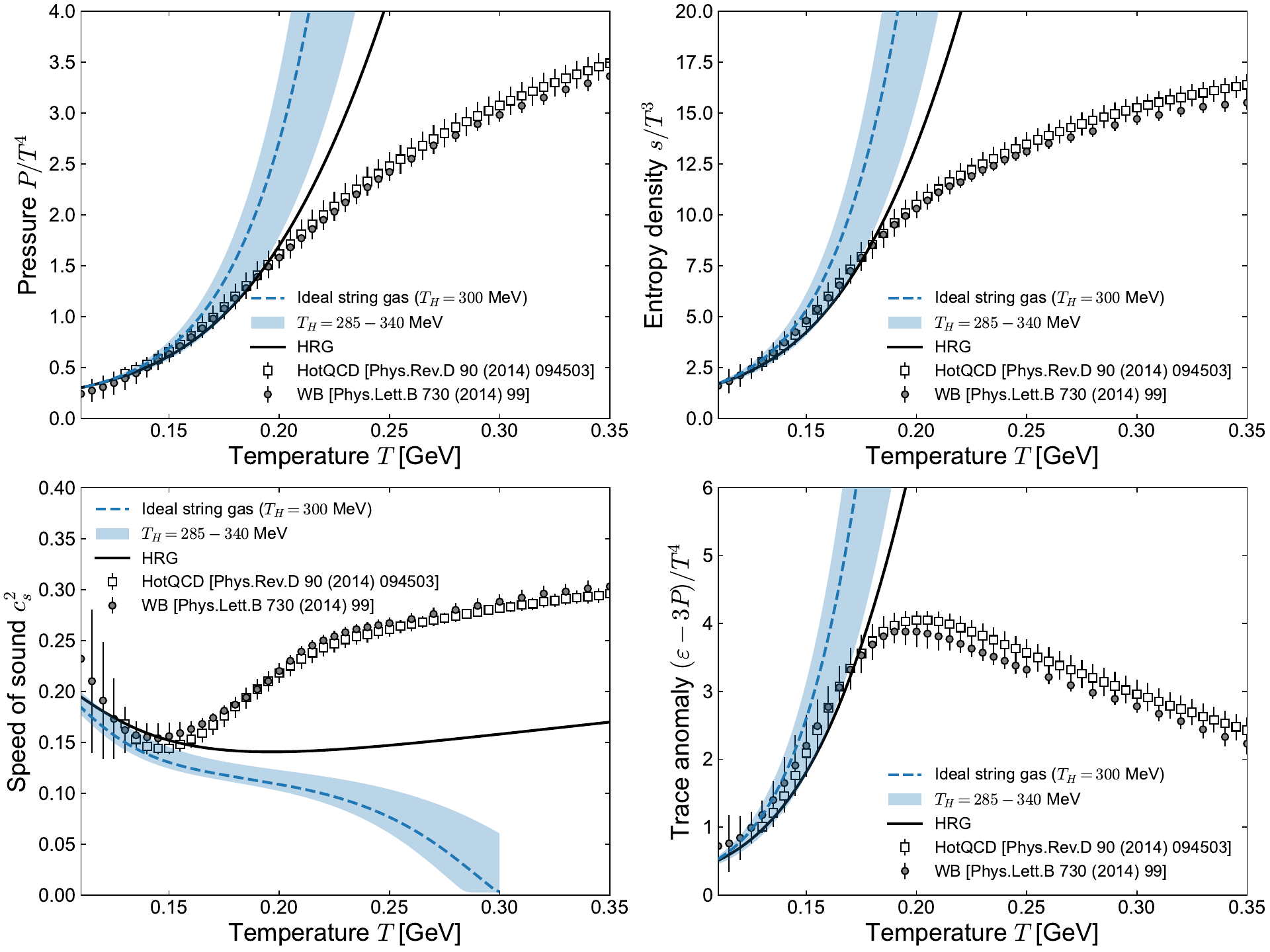}
    \caption{Bulk thermodynamic observables of the ideal string gas.  The dashed blue curves correspond to $\Th = 300 \,\text{MeV}$, and the shaded blue bands span $\Th = 285$--$340\,\text{MeV}$.
    The ideal HRG curves and lattice-QCD results from refs.~\cite{Borsanyi:2013bia,HotQCD:2014kol} are shown for comparison.
    \textit{Top left:} Pressure $P/T^4$.  \textit{Top right:} Entropy density $s/T^3$.  \textit{Bottom left:} Speed of sound $c_s^2$.  \textit{Bottom right:} Trace anomaly $(\varepsilon - 3P) / T^4$.
    }
    \label{fig:ideal}
\end{figure*}

Figure~\ref{fig:ideal} shows the bulk thermodynamic observables of the ideal string gas.
The dashed blue curves correspond to the representative value $\Th = 300\,\text{MeV}$, while the blue bands show the results over the physically motivated range $\Th = 285$--$340\,\text{MeV}$.
The lower endpoint, $\Th = 285 \,\text{MeV}$, is motivated by the critical temperature of the first-order deconfinement transition in pure Yang-Mills theory~\cite{Borsanyi:2022xml}.
It was also found that the Hagedorn temperature in the pure Yang-Mills glueball gas is close to the critical temperature~\cite{Meyer:2009tq}.
The upper endpoint, $\Th = 340 \,\text{MeV}$, is obtained from a fit to the observed hadron mass spectrum~\cite{Fujimoto:2025trl} and is also consistent with recent zero-temperature determinations of the string tension~\cite{Athenodorou:2020ani,Brambilla:2022het}.

At fixed $T$, the pressure, entropy density, and trace anomaly decrease as $\Th$ increases.
For these observables, the upper and lower edges of the band therefore correspond to $\Th=285$ and $340\,\mathrm{MeV}$, respectively.
The ordering is reversed for the speed of sound: its lower and upper edges correspond to $\Th=285$ and $340\,\mathrm{MeV}$, respectively.
Among the values shown, the larger values of $\Th$ give closer agreement with the ideal HRG and lattice QCD results, consistent with the best-fit value $\Th \simeq 340\,\text{MeV}$ obtained from the observed hadron spectrum.
The faster rise of the ideal string-gas pressure compared to the ideal-HRG result arises mainly because the continuum string spectrum extends to arbitrarily high masses, whereas the empirical hadron list used in the HRG calculation is truncated near $m\simeq 2\,\text{GeV}$.
The use of the asymptotic form~\eqref{eq:openstring} close to the channel thresholds may also contribute.

\subsection{Common excluded volume}

\begin{figure*}
    \centering
    \includegraphics[width=0.99\textwidth]{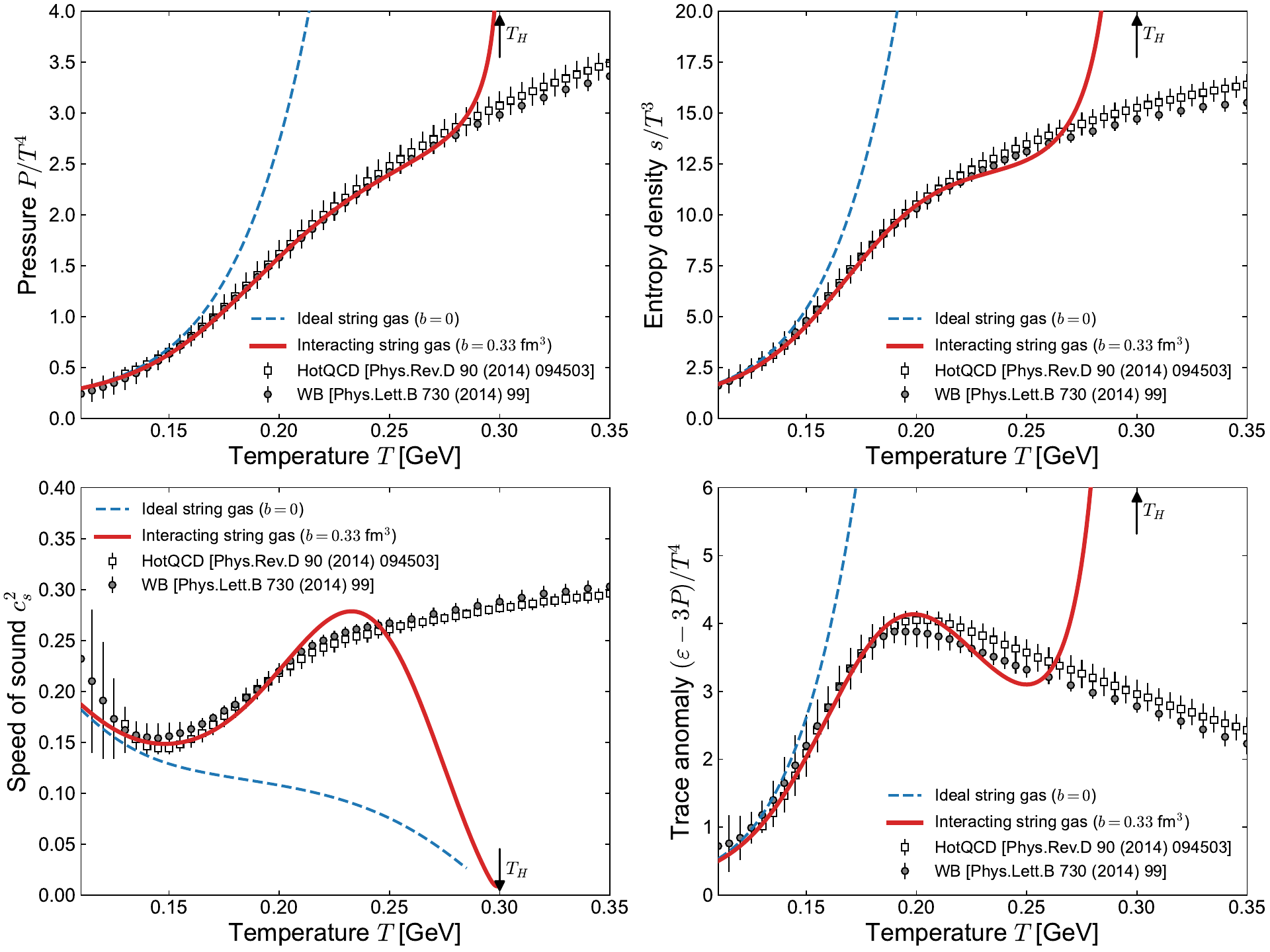}
    \caption{Thermodynamic quantities of the open string gas with $\Th = 300\,\text{MeV}$ incorporating the EV correction.
    The EV parameter $b$ is fitted to the temperature-pressure relation in the top left panel.
    \textit{Top left:} Pressure $P/T^4$.
    \textit{Top right:} Entropy density $s/T^3$.
    \textit{Bottom left:} Speed of sound $c_s^2$.
    \textit{Bottom right:} Trace anomaly $(\varepsilon - 3P) / T^4$.}
    \label{fig:evcommon1}
\end{figure*}

\begin{figure*}
    \centering
    \includegraphics[width=0.99\textwidth]{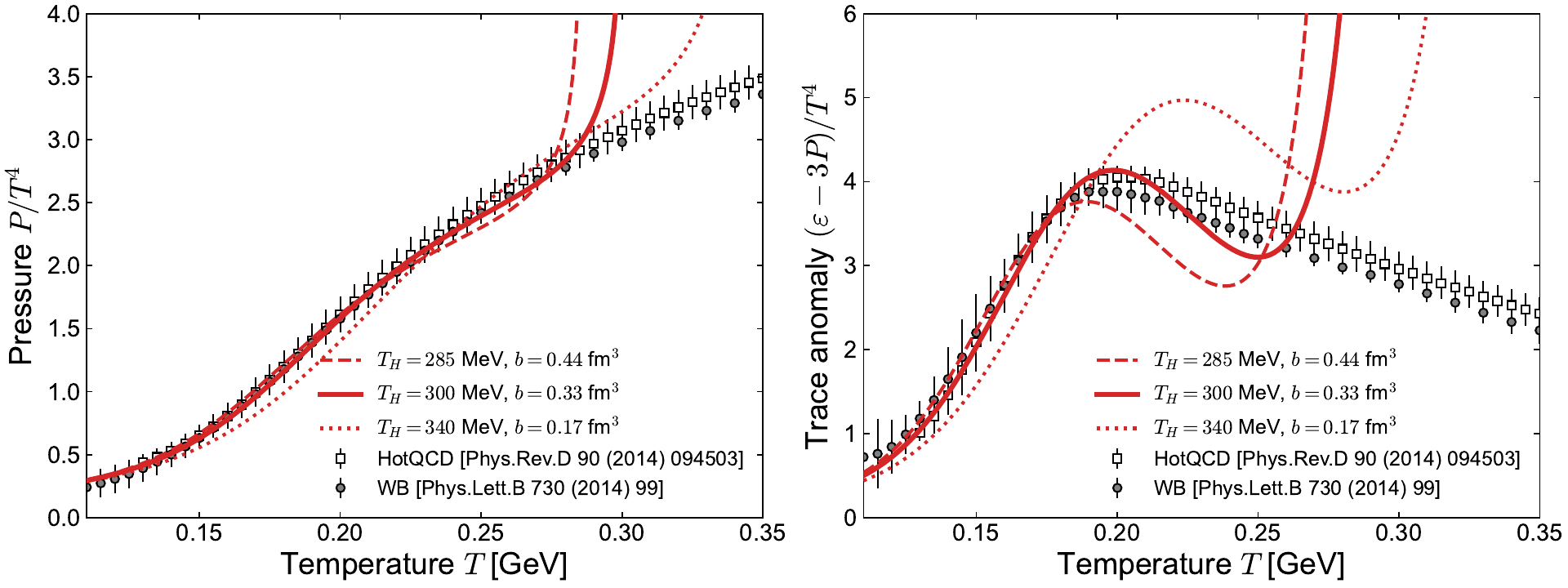}
    \caption{Thermodynamic quantities of open string gas incorporating the EV correction with varying $\Th$.
    The EV parameter $b$ is fitted to the temperature-pressure relation in the left panel for each $\Th$.
    \textit{Left:} Pressure $P/T^4$.
    \textit{Right:} Trace anomaly $(\varepsilon - 3P) / T^4$.}
    \label{fig:evcommon2}
\end{figure*}

We model residual repulsive interaction in the string-based HRG using a common EV parameter $b$.
Although such a simple EV prescription is not intended for precision observables such as the $\pi N$ phase shift in the $P_{33}$ channel~\cite{Lo:2017ldt}, it provides a minimal phenomenological framework for exploring string-string interaction in a thermal medium.
A microscopic treatment of string cutting, joining, and rearrangement lies beyond the scope of this work.
We therefore employ the EV prescription introduced in the previous section.

Figure~\ref{fig:evcommon1} shows the results for the interacting string gas at the representative value $\Th = 300\,\text{MeV}$.
As demonstrated in Fig.~\ref{fig:evcommon2}, the other values of $\Th$ produce qualitatively similar behavior, while $\Th = 300\,\text{MeV}$ gives the best quantitative agreement with the lattice-QCD results.

For each fixed value of $\Th$, we determine $b$ from an uncorrelated least-squares fit to the pressure data points from Ref.~\cite{Borsanyi:2013bia} in the range $110\leq T\leq280\,\mathrm{MeV}$, using the quoted pointwise uncertainties.
We note that lattice uncertainties at different temperatures are correlated~\cite{Hippert:2023bel,Shah:2024img}, but the precise correlation is unavailable in published form.
For this reason, we treat the extracted fit parameters as representative values, without assigning a fit uncertainty. 
For $\Th=300\,\mathrm{MeV}$, the fit gives $b\simeq0.33\,\mathrm{fm}^3$.
With the hard-sphere relation $b = (16 \pi / 3) r^3$, this corresponds to an effective radius $r\simeq 0.27\,\text{fm}$.
The associated packing fraction, $\eta = bn / 4$, is approximately $0.08$ at $T=200\,\text{MeV}$ and $0.16$ at $T= 250\,\text{MeV}$.
Thus, geometrical crowding remains moderate, although the EV correction becomes non-negligible toward the upper end of the temperature range.
As in the case of an ideal string gas, this model has a limiting temperature of $T = \Th$, where the thermodynamic quantities diverge due to the exponential density of string states.
A different choice of EV parametrization, such as a strong mass dependence of the EV parameter, could potentially mitigate this singularity at $T = \Th$~\cite{Ferroni:2008ej,Vovchenko:2018eod}.

The remaining panels in Fig.~\ref{fig:evcommon1} show the entropy density $s/T^3$, speed of sound $c_s^2$, and trace anomaly $(\varepsilon - 3P) / T^4$.
In particular, the model reproduces the peak structure in the trace anomaly well, whereas in the ideal string gas, it monotonically increases above $T_c$.
Similarly, the softest point in the speed of sound is also reproduced well, whereas in the ideal gas, it monotonically decreases.
These qualitative features of the EV effect have previously been explored in the HRG model~\cite{Vovchenko:2016rkn,Lysenko:2026ctu}.

Figure~\ref{fig:evcommon2} compares the results over the full range of $\Th$.
Fitting $b$ separately at each $\Th$ shows that smaller values of $\Th$ require larger excluded volumes.
The pressure remains well described throughout the range, and the trace-anomaly peak persists, although the quantitative agreement, especially for the latter, deteriorates as $\Th$ moves away from $300\,\text{MeV}$.

The improved thermodynamic description relative to HRG model descriptions can partly be attributed to the continuation of the string spectral density beyond the upper end of the empirical hadron list used in the HRG calculation.
The continuum extension should be interpreted as an effective spectral density, rather than as a prediction that all high-mass excitations appear as experimentally resolvable narrow resonances.
At high masses, broad and overlapping structures may contribute to thermodynamic observables without being identifiable as individual states.

\section{Conserved-charge fluctuations}

We now test whether the EV parameter determined from the pressure also describes conserved-charge fluctuations, and examine the effect of assigning different EV parameters to the mesonic and baryonic channels.
We define the conserved-charge susceptibilities as
\begin{align}
    \chi^{BQS}_{nml} &= \left. \frac{\partial^{n+m+l}[P(T,\boldsymbol{\mu})/T^4]}{\partial (\mu_B / T)^n \partial (\mu_Q/T)^m \partial (\mu_S/T)^l} \right|_{\boldsymbol{\mu}=0}\,.
\end{align}

\begin{figure*}
    \centering
    \includegraphics[width=0.99\textwidth]{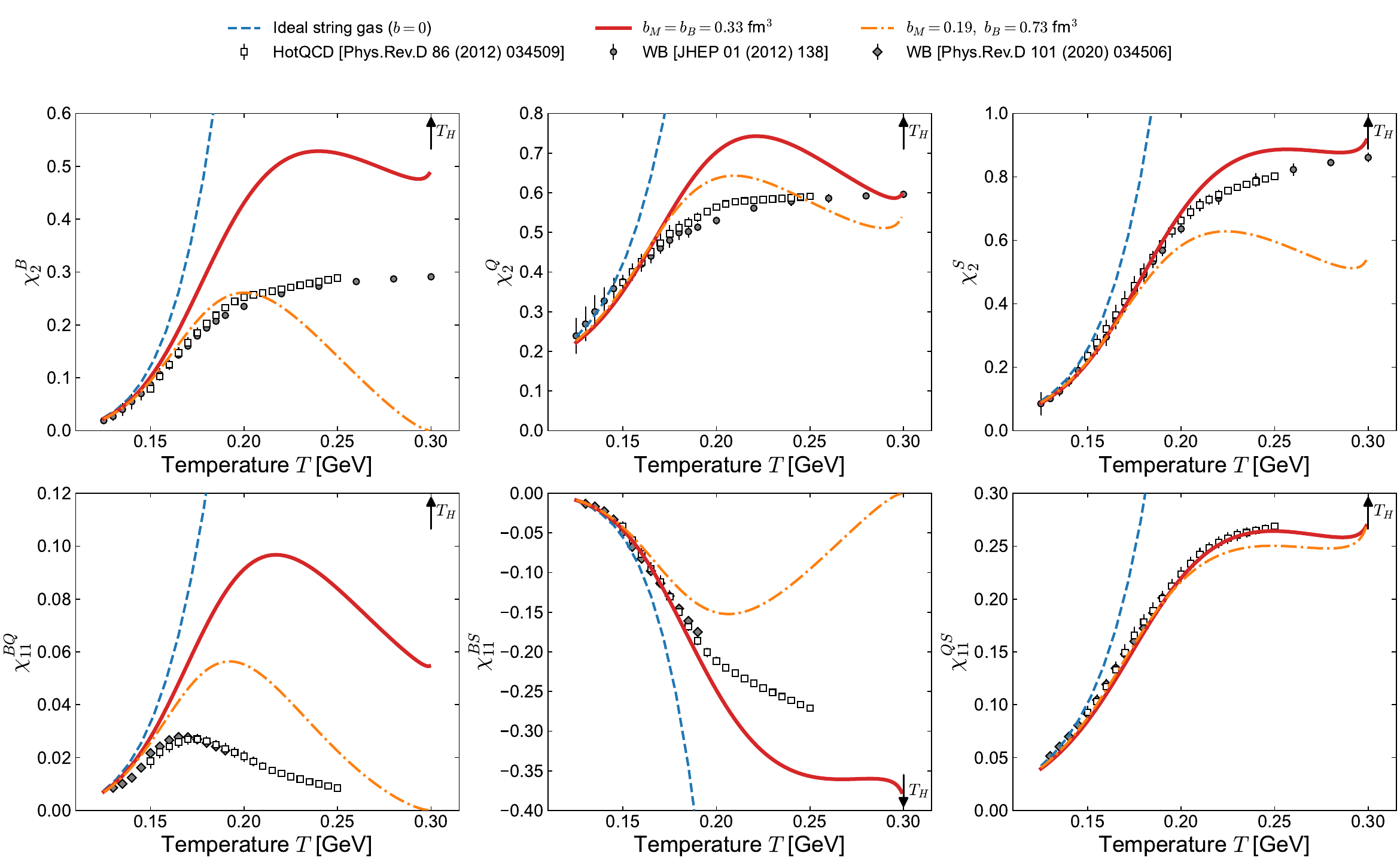}
    \caption{Second-order conserved-charge susceptibilities and correlations for the string-gas model with $\Th=300\,\mathrm{MeV}$. The dashed blue curves show the ideal string gas, the solid red curves the common-EV model with $b_M=b_B=0.33\,\mathrm{fm}^3$, and the dash-dotted orange curves the model with $b_M=0.19\,\mathrm{fm}^3$ and $b_B=0.73\,\mathrm{fm}^3$. Lattice-QCD results are shown for comparison. The top row displays $\chi_2^B$, $\chi_2^Q$, and $\chi_2^S$; the bottom row displays $\chi_{11}^{BQ}$, $\chi_{11}^{BS}$, and $\chi_{11}^{QS}$.}
    \label{fig:fluc}
\end{figure*}

Figure~\ref{fig:fluc} shows the diagonal second-order susceptibilities $\chi_2^B$, $\chi_2^Q$, and $\chi_2^S$ in the top row, and the mixed correlators $\chi_{11}^{BQ}$, $\chi_{11}^{BS}$, and $\chi_{11}^{QS}$ in the bottom row.
Throughout this section, we use $\Th = 300\,\text{MeV}$.

\subsection{Common excluded volume}

First, we employ the string model with the common EV parameter $b \simeq 0.33\,\text{fm}^3$, determined from the fit to the lattice QCD pressure in Fig.~\ref{fig:evcommon1}.
In Fig.~\ref{fig:fluc}, we compare our results with the lattice-QCD data from the Wuppertal-Budapest~\cite{Borsanyi:2011sw,Bellwied:2019pxh} and HotQCD~\cite{HotQCD:2012fhj} collaborations.
While $\chi_2^S$, $\chi_{11}^{BS}$, and $\chi_{11}^{QS}$ remain consistent with the lattice results up to $T \simeq 200\,\text{MeV}$, the baryon-sensitive observables $\chi_2^B$ and $\chi_{11}^{BQ}$ are substantially overestimated.
This indicates that the common EV parameter provides insufficient suppression in the baryonic sector and motivates a larger baryonic EV parameter.

Therefore, in the following subsection, we investigate whether the description can be improved by adopting different EV parameters for the meson and baryon sectors.

\subsection{Different excluded volumes for mesons and baryons}

\begin{figure*}
    \centering
    \includegraphics[width=0.99\textwidth]{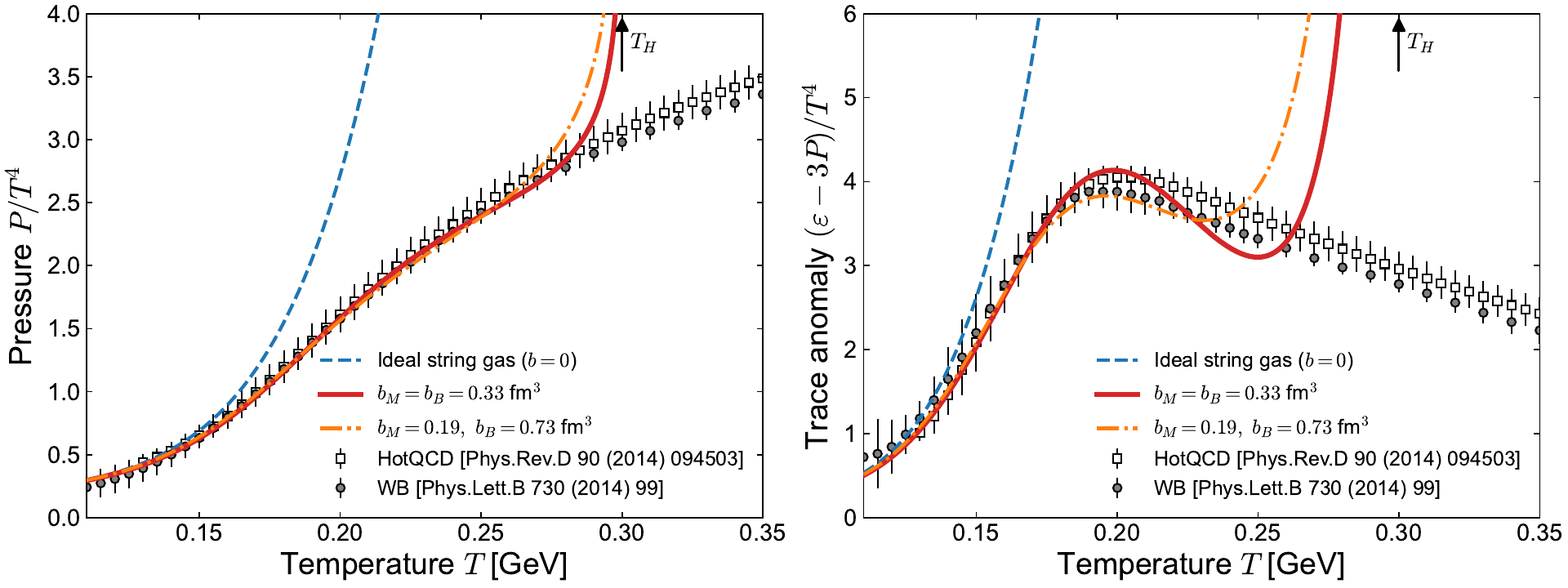}
    \caption{Bulk thermodynamic observables at $\Th=300\,\mathrm{MeV}$. The left and right panels show $P/T^4$ and $(\varepsilon-3P)/T^4$, respectively. The dashed blue curves denote the ideal string gas, the solid red curves the common-EV model with $b_M=b_B=0.33\,\mathrm{fm}^3$, and the dash-dotted orange curves the model with $b_M=0.19\,\mathrm{fm}^3$ and $b_B=0.73\,\mathrm{fm}^3$. Lattice-QCD results are shown for comparison.}
    \label{fig:thermo_evcross}
\end{figure*}

\begin{figure}
    \centering
    \includegraphics[width=0.98\columnwidth]{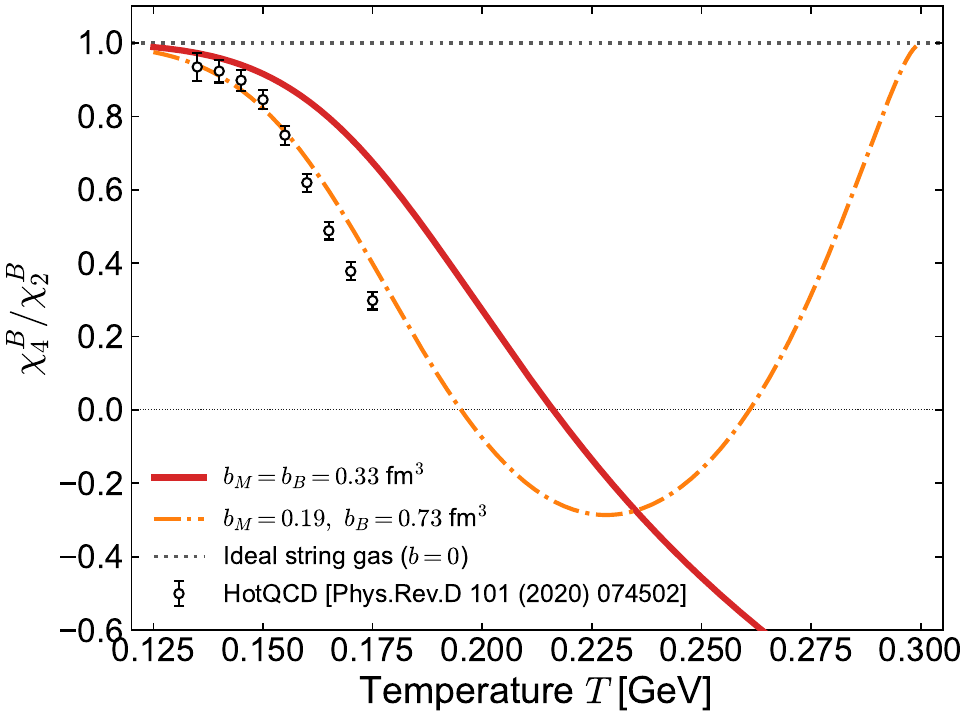}
    \caption{Fourth-to-second-order net-baryon susceptibility ratio $\chi_4^B/\chi_2^B$ at $\Th=300\,\mathrm{MeV}$. The solid red curve shows the common-EV result, the dash-dotted orange curve the result with different mesonic and baryonic EV parameters, and the dotted horizontal line the ideal-gas value. HotQCD data~\cite{Bazavov:2020bjn} are shown for comparison.}
    \label{fig:chi42}
\end{figure}

We next allow the mesonic and baryonic channels to have separate EV parameters, using the non-diagonal prescription defined in Eqs.~\eqref{eq:presmulti}--\eqref{eq:bMbB}.
At fixed $\Th=300\,\mathrm{MeV}$, we determine $b_M$ and $b_B$ from a simultaneous uncorrelated least-squares fit.
The fit includes $P/T^4$, $(\varepsilon-3P)/T^4$, $s/T^3$, and $c_s^2$ from Ref.~\cite{Borsanyi:2013bia} over $110\leq T\leq240\,\mathrm{MeV}$, together with the results for $\chi_2^B$, $\chi_2^S$, and $\chi_{11}^{QS}$ from Ref.~\cite{HotQCD:2012fhj} over $150\leq T\leq200\,\mathrm{MeV}$.
All pointwise residuals are normalized by the quoted uncertainties and included in the objective function with equal weight. 
Neglecting correlations among the data, the fit yields
$b_M=0.188\,\mathrm{fm}^3$ and $b_B=0.728\,\mathrm{fm}^3$.
As in the common-EV case, because correlations among the lattice observables and parameter uncertainties are not included, these values should be interpreted as effective calibration parameters rather than precise statistical determinations.

Figure~\ref{fig:thermo_evcross} verifies that introducing separate $b_M$ and $b_B$ does not change the qualitative bulk behavior.
In fact, the agreement with the lattice QCD data on the trace anomaly is arguably modestly improved.

Figure~\ref{fig:fluc} shows the effect of separate EV parameters for mesons and baryons on conserved charge susceptibilities.
As expected, the separate-EV prescription improves the fitted description of $\chi_2^B$ over the calibration range.
Among the observables not fixed by the calibration, it shifts $\chi_2^Q$
toward the lattice-QCD results over part of the temperature range, although the improvement is not uniform and quantitative agreement is not achieved.
As shown in Fig.~\ref{fig:chi42}, the additional EV degree of freedom also produces a stronger suppression of $\chi_4^B/\chi_2^B$, bringing its temperature dependence closer to the lattice results over the available data range.
These results indicate that distinguishing the mesonic and baryonic eigenvolumes captures part of the relevant sector dependence, but the improvement remains partial and
observable dependent.
In particular, previous analyses of second- and fourth-order mixed susceptibilities within the HRG model indicated reduced EV strength in the strangeness sector~\cite{Karthein:2021cmb,Vovchenko:2026ubv}. We leave such an analysis for future studies.

Noticeable tension remains in $\chi_{11}^{BQ}$, while the bulk pressure remains well described.
The sizable overestimate of $\chi_{11}^{BQ}$ indicates that the net charge-weighted baryonic response is too large in the present model.
In the isospin-symmetric limit, however, $\chi_{11}^{BQ}=(\chi_2^B+\chi_{11}^{BS})/2$, so the discrepancy may originate either from insufficient repulsion strength in the light-flavor baryon sector or from excessive repulsion in the strange-baryon sector.
The large degeneracy of the $\Delta$ channel makes it a plausible contributor, but a channel-resolved decomposition is required to establish this interpretation.

In addition to susceptibilities, lattice calculations at imaginary baryon chemical potential provide Fourier coefficients of the net-baryon density~\cite{Vovchenko:2017xad}.
In the ideal string gas, only the leading Fourier coefficient is non-zero, while lattice QCD indicates an alternating sign structure.
We have checked that below $T \simeq 200$~MeV, the interacting string gas generates the alternating-sign structure of the higher Fourier coefficients observed in lattice QCD.
At higher temperatures, however, the EV equations develop branch singularities in the complex-$\mu_B$ plane~\cite{Taradiy:2019taz,Oshima:2023bip},
whose location depends sensitively on the EV prescription. A detailed analysis of the resulting Fourier coefficients and analytic structure is therefore left for future work.

\section{Discussion and conclusions}

The present results show that an excluded-volume (EV) correction substantially improves the bulk thermodynamics of the ideal string gas.
For $\Th=300\,\text{MeV}$, a common EV parameter $b\simeq0.33\,\text{fm}^3$ reproduces the pressure and captures both the peak of the trace anomaly and the softest point of the speed of sound.
The corresponding packing fraction remains moderate over the temperature range of primary interest, but the EV contribution becomes increasingly important toward higher temperatures.

The EV parameter should not be interpreted uniquely as the geometrical excluded volume generated by a microscopic repulsive force between strings.
It may also encode string overlap and rearrangement, resonance broadening, modifications of the high-mass spectral density, and other many-body effects.
The fitted hierarchy $b_B>b_M$ therefore demonstrates that the baryonic contribution requires stronger effective suppression within the present model, but does not by itself establish a stronger microscopic baryon-baryon repulsive force.
Qualitatively, this hierarchy is consistent with prior HRG model analyses of susceptibilities in the chiral crossover temperature range.

The value $\Th\simeq300\,\text{MeV}$ preferred by the bulk thermodynamic comparison is lower than the value inferred from fits to the vacuum hadron spectrum.
This preference for $\Th\simeq300\,\text{MeV}$ should not be interpreted as a precise determination.
It is conditional on the temperature range included in the fit, since the high-temperature data are particularly sensitive to the rapid Hagedorn growth and its interplay with the EV correction.
Restricting the fit to lower temperatures would weaken, and may shift, the preferred value.
One possible interpretation is that the effective string tension or high-mass spectral density is modified in the thermal medium.
The present analysis does not determine such a medium dependence microscopically, however, because $\Th$ is treated as an external model parameter.
A direct comparison with finite-temperature determinations of the string tension would be required to establish this interpretation.

The success of the string-gas description above the pseudocritical temperature indicates that string-like or hadronic correlations remain thermodynamically relevant in this regime.
It does not, by itself, demonstrate that the system remains confined.
Rather, it suggests an intermediate region in which effective string-like and quark-gluon descriptions may both capture complementary aspects of the same thermodynamics.

Several systematic uncertainties remain.
These include the use of the asymptotic string density near the channel thresholds, the Maxwell-Boltzmann treatment of the light mesons, the assumption of mass-independent excluded volume parameters, the correlation between $\Th$ and the EV parameters, and the omission of covariance matrices in the combined calibration to lattice data.
Assessing these effects will be necessary for a quantitative determination of the effective string interactions.
These will also be relevant for the possible extension to non-zero baryon densities, which we leave for future work.

At asymptotically high temperatures, QCD thermodynamics should be described in terms of partonic degrees of freedom and weak-coupling methods~\cite{Kajantie:2002wa,Bresciani:2025vxw}.
The present string-gas model is intended for the intermediate-temperature regime and is not expected to remain applicable in this asymptotic region.
In the intermediate-temperature regime, one may expect a dual description of the same thermodynamics in terms of confined degrees of freedom, represented by strings in the current model, and partonic degrees of freedom, which need not be weakly coupled.
Whether calculations based on the string-gas model or weak-coupling methods are quantitatively controlled in this regime is, however, a separate question.
Establishing a quantitative matching between the string and perturbative descriptions is beyond the scope of this work.

In summary, an open-string description supplemented by excluded-volume corrections provides a compact model description of lattice QCD data on several thermodynamic observables in the intermediate temperature regime.
For $\Th=300\,\text{MeV}$, a common EV parameter $b\simeq0.33\,\text{fm}^3$ reproduces the bulk equation of state, while the conserved-charge susceptibilities favor a stronger effective suppression in the baryonic sector, with $b_M\simeq0.19\,\text{fm}^3$ and $b_B\simeq0.73\,\text{fm}^3$ in the present calibration.
The remaining discrepancies in the charge-sensitive observables show that a two-parameter EV prescription does not provide a complete description.
Nevertheless, the results support the relevance of string-like degrees of freedom above the pseudocritical temperature and motivate more systematic studies of the baryonic spectrum, medium-dependent string properties, and residual many-string correlations.

\begin{acknowledgments}
We thank Larry McLerran for discussions.
The work of YF is supported by JSPS KAKENHI Nos.~25K23388 and 26K17140. VV was supported by the U.S. Department of Energy, Office of Science, Office of Nuclear Physics, Early Career Research Program under Award No.~DE-SC0026065.
\end{acknowledgments}

\appendix
\section{Closed-form results for the common-eigenvolume model}

One can write the pressure with the EV correction in the common-eigenvolume model \eqref{eq:PEV} in closed form.
This is achieved by using the Lambert W-function, defined by the following equation for any complex number $z$:
\begin{equation}
    z = W(z) \exp[W(z)]\,.
\end{equation}
The pressure with the EV correction in the grand canonical ensemble \eqref{eq:PEV} precisely has this structure, namely,
\begin{equation}
    b \sum_i\phi_i(T)e^{\mu_i/T} = \frac{b P(T,\boldsymbol{\mu})}{T} \exp\left[\frac{b P(T,\boldsymbol{\mu})}{T}\right]\,.
\end{equation}
The solution of this transcendental equation can thus be written explicitly by using the Lambert W-function as~\cite{Noronha-Hostler:2012ycm, Taradiy:2019taz}
\begin{align}
    P(T,\boldsymbol{\mu}) = \frac{T}{b} W\left[b \sum_i\phi_i(T)e^{\mu_i/T}\right]\,.
\end{align}

For $n+m+l=2$, the diagonal and off-diagonal susceptibilities are
\begin{align}
    \chi^{BQS}_{nml}
    &=
    \left.
    \frac{
    \partial^{n+m+l}[P(T,\boldsymbol{\mu})/T^4]
    }{
    \partial(\mu_B/T)^n
    \partial(\mu_Q/T)^m
    \partial(\mu_S/T)^l
    }
    \right|_{\boldsymbol{\mu}=0}
    \\
    &=
    \frac{
    \left(\sum_i B_i^n Q_i^m S_i^l\phi_i\right)
    W_{\mathrm{tot}}
    }{
    bT^3\phi_{\mathrm{tot}}
    \left(1+W_{\mathrm{tot}}\right)
    }\,,
\end{align}
where $\phitot = \sum_i \phi_i$ and $W_{\mathrm{tot}} = W(b\phi_{\mathrm{tot}})$.
Similarly, the fourth-order susceptibility can be written as
\begin{align}
    \chi^B_4 &= \frac{\phiB \left[ (\phitot - 3 \phiB) \left(1 + W_{\mathrm{tot}}\right)^2 + 3 \phiB \right]  W_{\mathrm{tot}}}{bT^3 \phitot^2 \left(1 + W_{\mathrm{tot}}\right)^3}\,,
\end{align}
where $\phiB = \sum_i B_i^2 \phi_i$.

\bibliography{EV}
\end{document}